# Development of Slow Control Package for the Belle II Calorimeter Trigger System

C.H. Kim, S.H. Kim, I.S. Lee, H.E Cho, Y.J. Kim, J.K. Ahn, E.J. Jang, S.K. Choi,
Y. Unno and B.G. Cheon

*Abstract*—The Belle II experiment at the SuperKEKB e+e- collider in KEK, Japan does start physics data-taking from early of 2018 with primary physics goal that is to probe the New Physics effect using heavy quark and lepton weak decays. During trigger and DAQ operation upon beam collision, it is important that Belle II detector (Fig. 1) status have to be monitored in a process of data-taking against an unexpected situation. Slow control system, built in the Control System Studio (CSS) which is a GUI window design tool based on Eclipse, is one of monitoring and controlling systems in Belle II operation. Database and archiver servers are connected to slow control system. Experimental parameters are downloaded to Belle II main database server which is based on PostgreSQL. Real-time results are stored in archiver server which is based on EPICS (The Experimental Physics and Industrial Control System) archiver appliances and tomcat which is open-source java servlet container. In this study, we report the development of slow control system for the Belle II electromagnetic calorimeter (ECL) trigger system.

*Index Terms*— Belle II, Calorimeter, Trigger, Database, GUI, Slow Control.

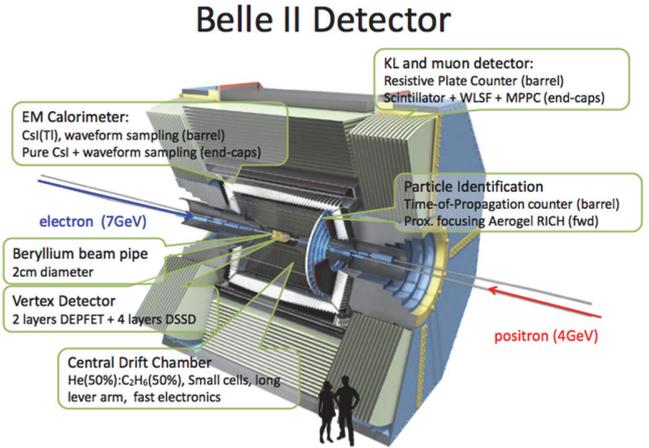

Fig. 1. The Belle II detector composed of five sub-detector components.

## I. Introduction

THE Belle collaboration have contributed to discover flavor structure of elementary particle, especially charge-parity(CP) violation of B meson decay in the summer of 2001. The Cabibbo-Kobayashi-Maskawa (CKM) mechanism, which gives explanation of the phenomenon, is experimentally confirmed by Belle collaboration [1]. M. Kobayashi and T. Maskawa were awarded the Nobel Prize 2008 by the confirmation [2]. Roughly, 1 ab$^{-1}$ data were collected by the experiment and it gives us clues to discover new physics if we have more data.

The Belle II detector in Fig.2 at the SuperKEKB collider is successor of Belle experiment at KEKB collider [3]. Target instantaneous luminosity of SuperKEKB is 40 times higher than KEKB [4]. Due to increase of luminosity, much high-level beam background is anticipated. For this reason, trigger system has been upgraded to have robustness and flexibility for preparing Phase III equipped with full detectors in 2019.

The level 1 hardware trigger for the Belle II operation is required to achieve high trigger efficiency from $\Upsilon(4S) \rightarrow B\bar{B}$ as 100%, 5 μs trigger latency and 30 kHz maximum trigger rate with keeping the trigger efficiencies of physics processes of interest as high as possible [1][4].

## II. Belle II Hardware Trigger System

A schematic of the Belle II hardware trigger systems is shown in Fig. 2. There are 4 sub-trigger systems, which are Central Drift Chamber (CDC) charged trigger, Barrel Particle Identification Detector (BPID) trigger, Electromagnetic Calorimeter (ECL) neutral trigger, and $K^0_L$ and μ (KLM) trigger. CDC trigger provides charged track information such as momentum, position, charge, multiplicity and so on. BPID

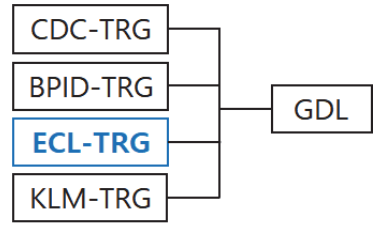

Fig. 2. Belle II Trigger system – Each sub-trigger outputs are delivered to Global Decision Logic (GDL)

This research was supported by Basic Science Research program through the National Research Foundation of Korea (NRF-2016K1A3A7A09005604, 2018R1A2B3003643) and by the research fund of Hanyang University (HY-2009).

C.H. Kim (primary author) and B. G. Cheon (corresponding author) are with the Department of Physics, the Research Institute for Natural Sciences, Hanyang University, Seoul, 04763 Korea (e-mail : chkim@hep.hanyang.ac.kr and bgcheon@hanyang.ac.kr).

S. H. Kim, I.S Lee, H.E Cho and Y. Unno are with the Department of Physics, Hanyang University, Seoul, 04763 Korea.

Y.J. Kim and J.K. Ahn are with the Department of Physics, Korea University, Seoul 02841 Korea.

E.J. Jang and S.K. Choi are with the Department of Physics, Gyeongsang National University, Jinju, 52828 Korea.

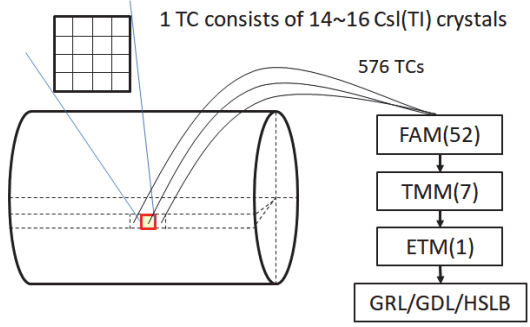

Fig. 3. ECL Trigger system – Signals from 576 TCs are processed on FAM, merged on TMM. Final signal generated on ETM and delivered to GRL, GDL and HSLB.

trigger provides precise timing and hit topology information. ECL trigger provides total energy, cluster information, and Bhabha identification information of electromagnetic particles. KLM trigger provides $K_L^0$ and μ track information. All the sub-trigger information is collected to Global Decision Logic (GDL) and final trigger decision is performed by this module. GDL output is delivered to the Belle II Data Acquisition system (DAQ) [4][5].

## III. ECL TRIGGER SYSTEM

Fig. 3 shows schematic of ECL Trigger system. The Trigger Cell (TC) which is composed of 4x4 CsI(*Tl*) scintillation crystals is a basic unit of the ECL trigger system and 576 TCs are installed in this system. FADC Analysis Module (FAM) receives all the TC outputs. After signal digitization on FAM modules, it provides TC energy and timing information to the Trigger Merger Module (TMM) via the GTX serial link. The TMM merges FAM outputs and sends it to the ECL Trigger Master (ETM). By piecing all the TC information together, ETM generates final ECL trigger outputs which are physics trigger and Bhabha trigger information [4].

## IV. ECL TRIGGER SLOW CONTROL SYSTEM

Slow control system, built in the Control System Studio (CSS) which is a GUI window design tool based on Eclipse, is one of monitoring and controlling systems in Belle II operation.

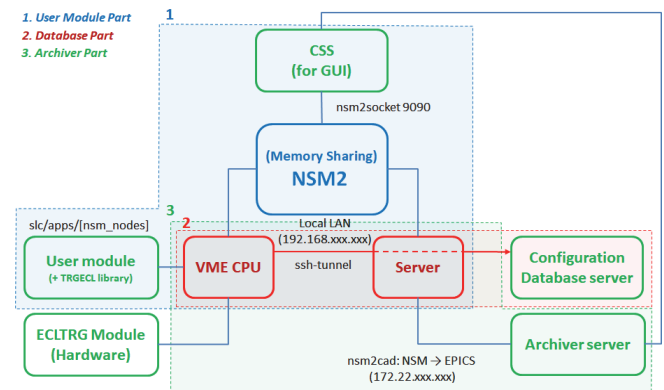

Fig. 4. Schematic of the ECL trigger slow control system. The system consists of 3 parts which are user module, database, and archiver part.

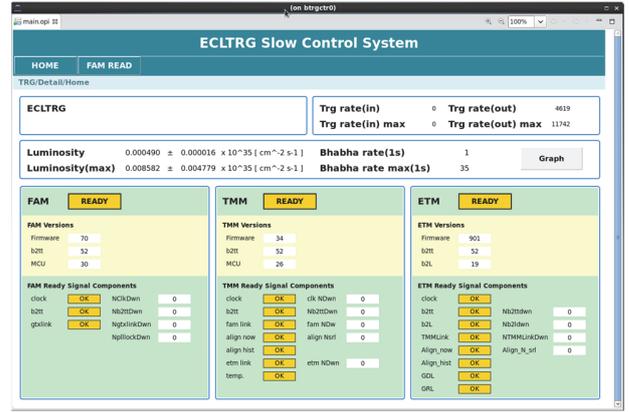

Fig. 5. User module – Readiness and firmware versions of each sub-system of ECL trigger are shown in user module.

Monitoring period is second order and this period is slow compared to readout speed of experimental data. Thus, the system is called slow control system. With this system, it is possible to monitor the detector status all the time, control the detector and handle an unexpected situation. ECL trigger library is successfully implemented in the slow control system, so it is possible to control ECL trigger with this system. The system consists of user module, database and archiver parts. Data and parameters are shared between them via Network Shared Memory for Belle 2 (NSM2) and Control System Studio(CSS) as shown in Fig. 4 [6].

### A. User module

User module part is GUI control panel, which is built by the CSS. The file extension is OPI (Operator Panel Interface). As the extension name means, the purpose of user module is to provide an intuitive control panel to an operator. ECLTRG monitoring panel is shown in Fig. 5. This user module shows status of each sub-system such as version and readiness. ECLTRG initialization and reset buttons are also prepared.

### B. Configuration database

Database part which is based on PostgreSQL package which is widely used for web server database. Connection between our slow control system and Belle II main database server is well established, so it is possible to download/upload experimental parameters to Belle II main database server via slow control system. A NSM2 node which automatically downloads ECL trigger parameters to the main database server by run number and time is made. Fig. 6 is schematic of how the nodes works. When each run starts, some of currently applied parameters on hardware such as firmware versions (In total 7 parameters which are FAM-FPGA/CPLD/MCU, TMM-FPGA/CPLD/MCU and ETM FPGA version), FAM energy thresholds (576) and timing offsets (576) are automatically saved on local machine, converted to appropriate format and downloaded to Belle II main database server by the node. After firmware update, fitter-CC (303264) will be added.

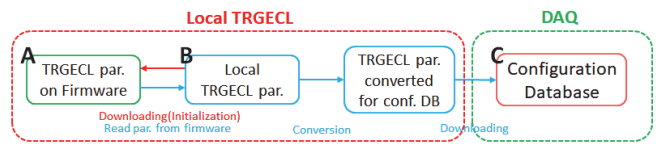

Fig. 6. Schematic of ECLTRG database part – When a run number is changed, a NSM node reads parameters from firmware (A), save the parameters on ECL trigger local server (B), convert them to appropriate format and finally download them to Belle II main configuration database server (C).

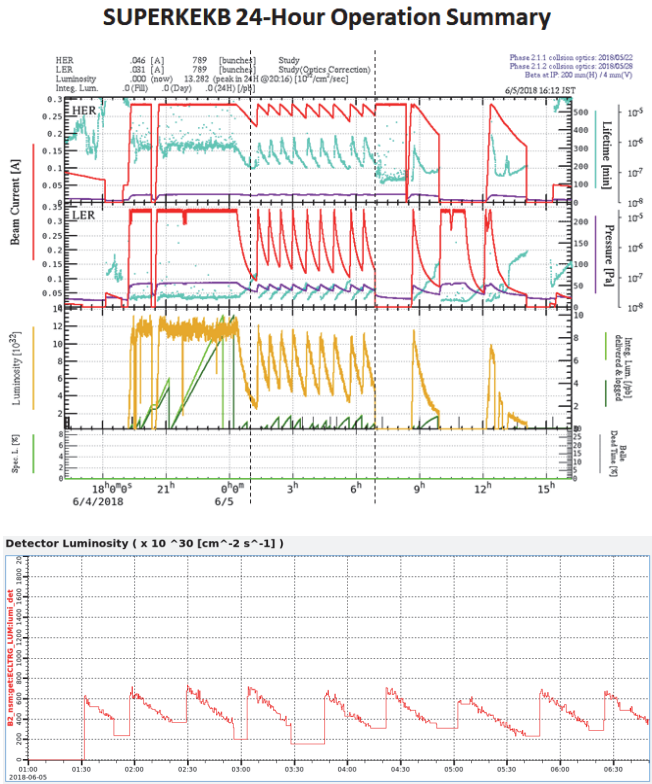

Fig. 7. Upper figure is SuperKEKB condition and lower figure is ECLTRG detector luminosity from ETM. Horizontal axis of both figures are time. Time range between two dashed lines on upper figure is same as full time range of lower figure. From the figures, we can see SuperKEKB condition and luminosity from our system is consistent.

*C. Archiver*

Archiver part based on EPICS (The Experimental Physics and Industrial Control System) archiver appliances and Apache Tomcat which is open-source java servlet container stores results of the experiment in real-time manner. To explain archiving process, experimental data from VME CPU is transferred to ECL trigger control server through NSM2, and the server converts NSM2 format to EPICS, finally EPICS format data is stored in Belle II main archiver server.

Around 100 PVs are currently archived. Archived data is not only for monitoring ECLTRG status, but also for SuperKEKB machine tuning and beam background study. SuperKEKB condition and ECL trigger detector luminosity are shown in Fig. 7 as an example. These figures let us know that shapes of ECL trigger detector luminosity and SuperKEKB providing luminosity are consistent each other.

V. CONCLUSION

1st version of slow control for ECL trigger system is prepared for stable Belle II operation. Versions and ready signal of each sub-triggers and initialization and reset command of whole system are included in user module. Some ELC trigger experimental parameters are automatically saved on Belle II main database server. Several EPICS PVs are archived such as luminosity, averaged TC hit-rate and so on. The archived PVs are not only for our system, but also helpful to SuperKEKB machine tuning.